\documentclass[a4paper,11pt]{article}
\pdfoutput=1 

\usepackage{jcappub} 

\usepackage[T1]{fontenc} 

\usepackage{xcolor}
\usepackage{amsmath}
\usepackage{natbib}
\usepackage{caption}
\usepackage{subcaption}

\usepackage{printlen}

\title{\boldmath Numerical Modeling of Time Dependent Diffusive Shock Acceleration}

\author[a,b,1]{S.~Aerdker,\note{Corresponding author.}}
\author[a, b]{L.~Merten,}
\author[a, b, c]{J.~Becker Tjus}
\author[a, b]{D.~Walter}
\author[a, b]{F.~Effenberger}
\author[a, b]{H.~Fichtner}

\affiliation[a]{Theoretical Physics IV, Plasma Astroparticle Physics, Faculty for Physics and Astronomy, Ruhr University Bochum, 44780 Bochum, Germany}
\affiliation[b]{Ruhr Astroparticle and Plasma Physics Center (RAPP Center), Germany}
\affiliation[c]{Department of Space, Earth and Environment, Chalmers University of Technology, 412 96 Gothenburg, Sweden}

\emailAdd{sophie.aerdker@rub.de}
\emailAdd{lukas.merten@rub.de}
\emailAdd{julia.tjus@rub.de}
\emailAdd{dw@tp4.rub.de}
\emailAdd{fe@tp4.rub.de}
\emailAdd{hf@tp4.rub.de}

\abstract{

Motivated by cosmic ray (CR) re-acceleration at a potential Galactic Wind Termination Shock (GWTS), we present a numerical model for time-dependent Diffusive Shock Acceleration (DSA). We use the stochastic differential equation solver (DiffusionSDE) of the cosmic ray propagation framework CRPropa3.2 with two modifications: An importance sampling module is introduced to improve statistics at high energies in order to keep the simulation time short. An adaptive time step is implemented in the DiffusionSDE module. This ensures to efficiently meet constraints on the time and diffusion step, which is crucial to obtain the correct shock spectra.
The time evolution of the spectrum at a one-dimensional planar shock is verified against the solution obtained by the grid-based solver VLUGR3 for both energy-independent and energy-dependent diffusion. 
We show that the injection of pre-accelerated particles can lead to a broken power law spectrum in momentum if the incoming spectrum of CRs is harder than the re-accelerated spectrum. If the injected spectrum is steeper, the shock spectrum dominates at all energies. 
We finally apply the developed model to the GWTS by considering a spherically symmetric shock, a spiral Galactic magnetic field, and anisotropic diffusion. The time-dependent spectrum at the shock is modeled as a basis for further studies. 

}

\begin{document}
\maketitle
\flushbottom

\section{Introduction}
\label{sec:intro}

The cosmic-ray energy spectrum and composition is by now well-studied from GeV up to ZeV energies. The origin of these highly energetic, charged particles is still largely unknown. Particularly unresolved is the question of the energy range between $10^{15}\,\mathrm{eV}$ and $3 \times 10^{18}\,\mathrm{eV}$. This energy range is defined by two kinks in the spectrum, the "knee" and the "ankle". Up to $\sim 10^{15}\,\mathrm{eV}$ it is believed that CRs are accelerated in the Galaxy, most likely at Supernova Remnants (SNRs) (see e.g.\ \citep{BeckerTjusMerten2020}). Above the ankle ($\sim 10^{18}\,\mathrm{eV}$) \citep{BeckerTjusMerten2020} CRs have gyroradii larger than the Galaxy and are clearly of extra-galactic origin. The breaks in the spectrum may indicate changes in the contributions to the spectrum: The spectral softening at the knee could be due to the maximal energy reached by Galactic sources. It could also result from a change in the energy-dependence of the residence time of CRs. The break at $\approx 200-300\,\mathrm{GV}$ especially at lower energies (\cite{BeckerTjusMerten2020} and references therein) are also attributed to a change in the transport properties.

In general, the observed power-law spectrum indicates that CRs are accelerated by stochastic processes. There are two possible scenarios of how to accelerate particles stochastically, both developed early-on by Fermi. Fermi Second-order Acceleration \citep{FermiII} was suggested first as acceleration on moving magnetized clouds in the Galaxy. Fermi First-order Acceleration \citep{AxfordEA77, Krymskii77, Bell78I, Bell78II, BlandfordOstriker78} is more efficient: CRs are scattered on both sides of a shock and may cross the shock front repeatedly. Each time they cross the shock, they are accelerated depending on their energy, $\Delta E \propto E$. The energy gain can be derived by Lorentz boosts between the upstream and downstream reference frames. For non-relativistic, stationary shocks and neglecting non-linear interaction between the CRs and the background plasma the slope of the stationary shock spectrum is purely dependent on the shock's compression ratio. Even for non-linear diffusion coefficients the spectral form is unchanged \citep{WalterEA2022}. 

 There are some limitations to DSA, one of which is that particles must be able to reach the shock from downstream to upstream again, see, e.g. \cite{KallenbachEA2005}. In order to cross the shock, CRs must have been accelerated to energies that give them a gyroradius larger than the shock width, which is also known as the injection problem. Also, CRs have to be contained in the acceleration region to experience ample acceleration. This implies a diffusive particle transport with diffusion low enough that CRs are not escaping too quickly. Containment can also be implied by the geometry of the system, e.g. in stellar clusters particles can only escape downstream.
 
 An upper bound of the maximum energy produced by a source can be estimated by the Hillas criterion and depends on the particles' rigidity and the size of the acceleration region \citep{Hillas84}. Often magnetic field amplification due to CRs determines the maximum energy reached by DSA at the shock. There are other limiting factors like the lifetime of the accelerator or loss processes like synchrotron radiation. 
 
In the transition region from Galactic to extragalactic CRs the Galactic Wind Termination Shock (GWTS) might contribute to the sources of CRs via re-acceleration \citep{ThoudamEA2016, MertenEA2018}. A possible Galactic wind can be driven by, e.g., radiation or the CRs themselves (e.g.\ \citep{BreitschwerdtEA91, EverettEA2008, UhligEA2012, BustardEA2016, MaoEA2018}). It may form a shock when the supersonic wind slows down due to interaction with the Intergalactic Medium (IGM). Especially for starburst galaxies there is evidence for supersonic outflowing winds (e.g. \citep{LopezEA2017, VeilleuxEA2005}). 

CRs with their origin in the Galactic disk are advected outwards and are accelerated to higher energies at the GWTS. Either they leave the Galaxy or a fraction of the high-energy CRs may be able to propagate back to the Galaxy against the Galactic wind. The idea of particle acceleration at the GWTS was already discussed e.g.\ by \cite{JokipiiMorfill87}, \cite{ThoudamEA2016}, \cite{MertenEA2018}, and most recently by \cite{MukhopadhyayEtAl2023}. 
With the assumption that about two solar masses per year are advected outwards, the GWTS cannot be supported by the wind longer than $100\,\mathrm{Myr}$ \citep{BustardEtAl2017}. The finite lifetime of the shock may also impact the shock spectrum and its contribution to the observed cosmic-ray spectrum on Earth. 

Not only the origin of CR but also their transport properties change in the transition region. Particles undergo a random walk due to scattering on magnetic field turbulence. The interaction of CRs with the magnetic field turbulence depends on the gyroradius and the turbulence spectrum of the magnetic field. CRs with energies above $\sim 10^{18}\, \mathrm{eV}$ engage less in interaction with the turbulent Galactic magnetic field and can only be described by the diffusion approximation at late times, and with that, on larger scales \citep{KaapaEA2022}. The transition time from ballistic to diffusive transport generally grows with energy. In energy, it is still unclear where exactly the transitions between diffusive and ballistic transport as well as Galactic and extra-galactic CRs lie. 

Methods for the simulations of CR transport have to reflect that transition: For high-energy extra-galactic CRs the equation of motion is integrated. In this ballistic regime, individual particle trajectories in arbitrary magnetic field configurations can be simulated precisely. However, for Galactic CRs with energies below $10^{16}\,\mathrm{eV}$ this approach becomes computationally costly as most of the computation is spend on resolving the particles' gyration. Therefore, for Galactic CRs the transport equation is commonly used to describe diffusive transport in space and momentum. 

The simplified, assuming isotropic (in momentum space) CR distributions, Fokker-Planck equation, often called Parker transport equation (e.g.\ \citep{Schlickeiser2002})
\begin{equation}
\label{eq:TransportEq}
    \frac{\partial \mathcal{N}}{\partial t} + \vec{u} \cdot \nabla \mathcal{N} = \nabla \cdot \left( \hat{\kappa} \nabla \mathcal{N} \right) + \frac{1}{p^2} \frac{\partial}{\partial p} \left(p^2 D \frac{\partial \mathcal{N}}{\partial p} \right) + \frac{1}{3} \left( \nabla \cdot \vec{u} \right) \frac{\partial \mathcal{N}}{\partial \ln p} +  S(\vec{x}, p, t),
\end{equation}
describes the time evolution of the cosmic-ray differential number density $\mathcal{N} = p^2 f(\vec{x}, p, t)$ in space $\vec{x}$ and momentum $p = |\vec{p}|$, with $f$ being the CR distribution function. Individual terms describe advection of CR with a background flow $\vec{u}$, spatial diffusion described by the diffusion tensor $\hat{\kappa}$, momentum diffusion described by the coefficient $D$, adiabatic energy changes and cosmic-ray sources $S$.

In the test-particle picture, the average change in the particle's momentum can be calculated from changing the local fluid frames when the shock is crossed. Together with the probability to escape downstream and never return to the shock, the well-known power-law in momentum can be derived (e.g. \citep{Bell78I} or \textit{microscopic derivation} in \citep{Drury}). In the diffusive picture, the ensemble-averaged distribution function $f$ or density $\mathcal{N}$, are described by the transport equation in the frame in which the shock front is stationary. Here, diffusive shock acceleration results from adiabatic heating and the interplay between advection and spatial diffusion (e.g. \citep{Krymskii77}, \citep{KirkSchlickeiser88} or \textit{macroscopic approach} in \cite{Drury}).

The spatial diffusion tensor is often defined in a local coordinate frame, describing diffusion parallel and perpendicular to the local magnetic field. The momentum diffusion coefficient describes acceleration of CR in the presence of magnetic field turbulence (Second-order Fermi acceleration). The knowledge of both, spatial and momentum diffusion tensors, is essential for the complete description of CR transport. 

There is not yet a closed description of the diffusion tensor. Depending on the assumed magnetic field, coherent background and turbulent component, and the CRs properties analytical or numerical approximations might be available (e.g. \citep{DeMarcoEA2007, SnodinEA2016, GiacintiEA2018, Mertsch2020, ReichherzerEA2020, ReichherzerEA2022, KuhlenEA2022}). On the other, hand it can also be used as a free parameter and fitted to match, e.g., the observed primary and secondary ratios in a model of Galactic cosmic-ray transport. Since all of these approaches come with their own caveats, we will use normalized diffusion coefficients $\tilde{\kappa}_\parallel = 1$ throughout this work.

Considering the GWTS mentioned at the beginning, re-accelerated CRs are expected to be in the transition from diffusive to ballistic transport. We expect a finite lifetime for the GWTS, so that the transition time between the transport regimes is important. Close to the shock due to strong turbulence induced by the CRs themselves the diffusive approach is valid for higher energies than in the interstellar or IGM at a fixed time \citep{Bell78I, AmatoBlasi2006}. Particles escaping the acceleration region, however, cannot necessarily be described by the diffusion approximation depending on their energy. In order to assess the contribution of the GWTS to the cosmic-ray spectrum they must be propagated back to the Galaxy using either the ballistic or diffusive approach depending on their energy. Thus, the spectrum observed on Earth is further modulated by the energy-dependence and spatial variability of the diffusive transport. 

The CR propagation framework CRPropa3.2 \citep{CRPropa3.2} offers the possibility to simulate acceleration and propagation in both regimes. Also, the structure of CRPropa3.2, which is described in more depth in Section \ref{sec:Methods}, makes it relatively easy to define arbitrary magnetic field and shock configurations in up to three dimensions. Together with the GWTS scenario, this motivates an in-depth study of DSA using the diffusion approach of CRPropa3.2 \citep{MertenEA2017}. In test scenarios we explore different questions related to the re-acceleration of CR at the GWTS: How does the spectrum change when a finite shock lifetime is considered? What are the effects on the spectrum when CRs are already pre-accelerated to a power-law? To what extent does a finite acceleration region affect the maximal energy that can be reached?

First, we give a short overview of CRPropa3.2 and its diffusion approach which is based on Stochastic Differential Equations (SDEs) in Section \ref{sec:Methods}. A new module is presented that was specifically implemented to enhance statistics when simulating DSA with CRPropa3.2. 

In Section \ref{sec:1D} we show that DSA simulations with CRPropa3.2 result in the expected spectral slope for acceleration at one-dimensional planar shocks. We validate the obtained stationary spectra with predictions from theory and other ensemble-averaged approaches to DSA. The time evolution of the spectrum at the shock is compared to simulations that integrate the transport equation using the finite difference code VLUGR3 \citep{VLUGI, VLUGII, WalterEA2022}. We clarify constraints for simulating DSA based on SDEs. 
Based on those findings, we consider a finite shock lifetime, energy-dependent diffusion and the injection of pre-accelerated spectra. The effects of each modification on the spectrum at the shock or the downstream boundary are analysed separately.

In Section \ref{sec:2D} we take anisotropic diffusion into account and consider oblique shocks. Finally, we present the time-dependent spectrum at a simple model for the GWTS: A spherical symmetric shock and spiral magnetic field.

\section{Methods}
\label{sec:Methods}

In this Section we briefly introduce SDEs and how they relate to the transport equation (\ref{eq:TransportEq}). We explain how they are used in CRPropa3.2 to solve the transport equation. For the simulation of DSA a new module, \texttt{CandidateSplitting}, was implemented to enhance statistics at high energies.

    \subsection{Stochastic Differential Equations}
    
    One way to simulate DSA is the solution of the transport equation (\ref{eq:TransportEq}), a partial differential equation in reduced phase-space. The solution directly results in the distribution function of CRs. We followed this approach using the well-established VLUGR3 \citep{VLUGI, VLUGII} code to cross check our results. Another way is to make use of the connection between Fokker-Planck type equations and SDEs. In general, Fokker-Planck type equations describe the transition probability from one state to a set of other states. SDEs describe dynamical systems which are subject to noise. 
    
    In the following we discuss briefly the set of SDEs implemented in CRPropa3.2. For an in-depth explanation the reader is referred to \cite{MertenEA2017} or e.g.\ \citep{KoppEA2012}.

    In the absence of second-order Fermi acceleration processes, i.e.\ $D = 0$, and spatially constant diffusion the system of SDEs equivalent to the transport equation (\ref{eq:TransportEq}) is given by
    \begin{align}
    \label{eq:SDEI}
        \mathrm{d}\vec{x} &=  \vec{u}(x)\,\mathrm{d}t + \hat{B} \,\mathrm{d} \vec{\omega}_{t},\\
        \mathrm{d}p &= - \frac{p}{3} \nabla \cdot \vec{u} \, \mathrm{d}t,
        \label{eq:SDEII}
    \end{align}
    where $\mathrm{d} \vec{\omega}_{t} = \sqrt{dt} \, \vec{\eta}$ is a Wiener process with 
    $\eta_{\mathrm{i}}$ being random numbers from a unit normal distribution.
    The tensor $\hat{B}$ corresponds to spatial diffusion. In the local frame of the magnetic field line the diffusion tensor becomes diagonal. Drift terms due to curvature of the magnetic field lines are described by off-diagonal elements and are neglected in the following.
    With this assumption $\hat{B}$ is given by $B_{\mathrm{ij}} = \delta_{\mathrm{ij}} \sqrt{2 \kappa_{\mathrm{ij}}}$. 
    Equation (\ref{eq:SDEI}) describes the spatial displacement due to advection and stochastic fluctuations. Equation (\ref{eq:SDEII}) describes the adiabatic energy change. Without diffusion in momentum space it is not an SDE but an ordinary differential equation.  

    SDEs can also be written in integral form by making use of the It\^{o} integral \citep{Ito51}. In this form they can be approximated numerically (see e.g.\ \citep{KloedenPlaten92, Gardiner2009}).
    The SDE approach has the advantage that it is easy to implement and to extend to arbitrary geometries. We describe the first-order Euler-Maruyama scheme implemented in CRPropa3.2 for numerical solution of the SDE (Eq.\ (\ref{eq:SDEI})) in the following Section.

    \subsection{Simulation with CRPropa3.2}

    CRPropa3.2 has a modular structure with the \texttt{Candidate} class as central element. When the EoM is used, the candidate simply corresponds to the propagated particle. When the diffusive transport is used, the candidate corresponds to a phase-space element that is propagated. The phase-space element is also called \textit{pseudo-particle} in the following. The candidate module holds all information about the (pseudo-)particle that can be used and altered by other modules in the \texttt{ModuleList} that are successively called each simulation step. The user can add different kinds of modules to the \texttt{ModuleList}: Propagation, Acceleration, Interaction, Boundaries and Observers. Other modules handle the simulation environment (\texttt{MagneticField}, \texttt{AdvectionField}), the injection of candidates (\texttt{Source}) and the output. The modular structure makes it easy to set up simulations for various scenarios and to add new modules. 

    To model DSA, we use the \texttt{DiffusionSDE} module \citep{MertenEA2017}. Pseudo-particles are propagated parallel and perpendicular to a magnetic field, depending on the diffusion tensor 
    \begin{equation}
    \label{eq:kappa}
        \hat{\kappa} = \kappa_{\parallel} \begin{pmatrix}
            \epsilon & 0 & 0  \\
             0 & \epsilon & 0 \\
             0 & 0 & 1 
        \end{pmatrix} \left( \frac{E}{E_0} \right)^{\alpha},
    \end{equation}
    which is constant in space and can depend on energy. The diffusion coefficients for the normal and binormal direction of the magnetic background field line can be assumed to be the same for large curvature radii of the magnetic field lines $\kappa_\mathrm{n}=\kappa_\mathrm{b} \equiv \kappa_{\perp} = \epsilon \kappa_{\parallel}$.
    In Section \ref{sec:1DEnergy} energy-dependent diffusion is considered. Other than that, diffusion is considered to be energy-independent. 
    The SDE \ref{eq:SDEI} is integrated using the Euler-Maruyama Scheme (e.g.\ \citep{KloedenPlaten92})
    \begin{equation}
    \label{eq:EMScheme}
        \vec{x}_{n+1}-\vec{x}_n= \vec{u} \Delta t + \sqrt{2\hat{\kappa}}\Delta \vec{\omega}_n,
    \end{equation}
    with time step $\Delta t$ and random variables $\Delta \vec{\omega}_n = \vec{\omega}_{t_{n+1}} - \vec{\omega}_{t_n}$ drawn from a normal distribution with an expected value of zero and variance of $\Delta t$. 
    SDE methods have no constraints on the used time step. However, the choice of time step can be crucial to obtain correct results. See \cite{StraussEffenberger2017} for a discussion.
    
    Eq.\ (\ref{eq:EMScheme}) in general is defined in the lab frame. Since the diffusion coefficient in Eq.\ (\ref{eq:kappa}) is defined in the local magnetic field frame, the diffusive step $\sqrt{2\hat{\kappa}}\cdot \Delta \vec{\omega}_n$ is calculated in the orthonormal base of the magnetic field by integrating along the magnetic field line for parallel diffusion, calculating the perpendicular diffusive step and transforming back to the lab frame in each time step. 
    
    To ensure one-dimensional propagation in Section \ref{sec:1D} the magnetic field is set constant in x-direction and diffusion is allowed only parallel to the magnetic field, i.e. $\epsilon = 0$. In Section \ref{sec:2D} perpendicular diffusion is taken into account as well as a more complex magnetic field. 

     Acceleration is handled by the \texttt{AdiabaticCooling} module according to the momentum equation (Eq.\ (\ref{eq:SDEII})) based on the divergence of the shock profile specified in the \texttt{Advection\-Field} class. The advection field and magnetic fields used for simulations are specified in the following sections.

     Free-escape boundaries $L_{\pm}$ can be specified in the upstream and downstream region. Observer modules detect pseudo-particles when they cross those boundaries or the \texttt{Time{\-}Evo{\-}lution{\-}Observer} is used to detect pseudo-particles' positions and energies at given times $t_i$ during the simulation in the acceleration region. 

    \subsection{Analysis}

    The energy spectrum $J(E) = \mathrm{d}N/\mathrm{d}E$ of particles in the acceleration region is approximated by the histogram $\Delta N/\Delta E$, with $\Delta N$ being the number of pseudo-particles in each energy bin $\Delta E$. In the high-relativistic limit, assuming $E = p/\mathrm{c}$, $J \propto p^2 f(p,t)$. Since we expect a power-law spectrum, energy is binned with equal distance in logarithmic space. The error for each bin is then given by $\Delta J = J/\sqrt{\Delta N}$, any additional errors (e.g. from summing over time "snaps" as explained in Section \ref{sec:1DStationary}) are not included. The spectrum can be weighted by $E^2$ to highlight the spectral slope $s= -2$ predicted for acceleration at strong shocks. 
    
    \subsection{Candidate Splitting}

    With the assumptions of non-relativistic shocks and neglecting non-linear interactions, DSA produces a power-law spectrum in energy and the spectral slope depends only on the shock compression ratio. When mono-energetic candidates are injected at the shock and accelerated until they leave the acceleration region, the number of candidates decreases with increasing energy. Especially for less efficient acceleration at shocks with a low compression ratio, statistics at high energies may be so bad that it is difficult to evaluate whether the stationary solution is already reached or to determine the resulting spectral slope. 
    
    The number of injected candidates can be increased in order to get sufficient statistics at high energies, which comes at large computational costs. Another way to overcome this problem is to split candidates in $n_{\mathrm{split}}$ copies once they cross energy boundaries in log-space. For DSA the optimal splitting number depends on the expected spectral slope to compensate for the loss of candidates at higher energies. Each splitted candidate is assigned with a weight $w = 1/n_{\mathrm{split}}$ during the simulation. In the later analysis candidates are weighted accordingly to obtain the correct spectra. In order to determine the error of the spectrum for each energy bin all candidates can be used, which reduces the uncertainty at high energies.
    
\begin{figure}[tbp] 
        \centering 
        \includegraphics[width=.49\textwidth]{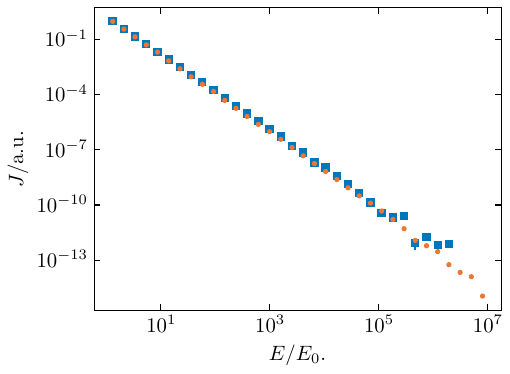}
        \hfill
        \includegraphics[width=.49\textwidth]{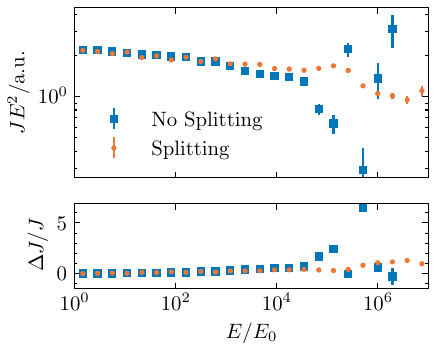}
        \caption{Left: Spectra obtained with and without using the \texttt{CandidateSplitting} module at simulation time $t = 400\;t_0$ evaluated at the shock, $x = [0,2]\,x_0$. The shock compression ratio is $q=4$ and both simulations reproduce the expected spectral slope $s=-2$. The simulation with candidate splitting reaches higher energies up to $10^7\;E_0$ with smaller errors. Right: Weighted spectra (top) and relative error $(J_{-2} - J)/J$ where $J_{-2}$ is the predicted spectrum (bottom). }
        \label{fig:CandidateSplitting}
    \end{figure}

    In Fig.\ \ref{fig:CandidateSplitting} resulting shock spectra simulated with and without candidate splitting are compared at the same simulation time ($t=400\,t_0$). The injected number of candidates is $N_0 = 10^6$ for the simulation without and $N_0 = 10^4$ with candidate splitting. For the latter, the actual number of simulated candidates increases during the simulation. Still, computational time goes down by a factor of $25$\footnote{CPU Time (Apple M1 Pro) for $N_0 = 10^4$ candidates with \texttt{CandidateSplitting}: $194\pm7.92\,\mathrm{s}$. CPU Time for $N_0 = 10^6$ candidates without \texttt{CandidateSplitting}: $5212\pm75\,\mathrm{s}$. Both averaged over 8 runs.} when candidate splitting is used, since fewer candidates are simulated in low-energy bins. For the set-up shown in Fig.\ \ref{fig:CandidateSplitting}, at $t = 400\;t_0$, the number of candidates is about $2.1 N_0$.

    Note that the depicted errors are given by $\Delta J = J/\sqrt{\Delta N}$. To approximate the stationary solution of the energy spectrum at the shock, energy of pseudo-particles stored during simulation are summed over time as explained in Section \ref{sec:1DStationary}. Such time related errors are not shown in Fig.\ \ref{fig:CandidateSplitting} and following figures.
    
    Splitting of candidates at energy bins in logarithmic space is one way of importance sampling used also in other approaches to DSA \citep{AchterbergSchure2010}. Another way is the introduction of artificial drift terms pushing particles to the shock \citep{StraussEffenberger2017, Zhang2000}. Such drift terms need to be extracted in the analysis. The candidate splitting method is, however, easier to apply to more complex configurations of the magnetic field and advection field. 

    Candidate splitting is implemented as an independent module in CRPropa3.2. It is possible to define individual energy bins and number of splits. For use in diffusive shock acceleration only the minimal and maximal energy as well as the expected spectral index need to be specified. 

\section{One-dimensional Diffusive Shock Acceleration}
\label{sec:1D}

In the ensemble averaged picture, energy gain at shocks is described by Eq.\ (\ref{eq:SDEII}). To apply this adiabatic description a finite shock width is considered since the advective speed $u(x)$ must be continuously differentiable. Deviations from a sharp shock transition can occur if a population of energetic particles is interacting with the thermal background plasma. In such case the latter can experience a decelerating force due to the pressure gradient resulting from the (energy density) distribution of the suprathermal particle population and either a so-called subshock forms or the shock weakens into a smooth transition (see, e.g., \citep{HaggertyEA2020, MalkovEA2010, leRouxEA2000, Ko95, LeeAxford88, DruryVoelk81}).
The one-dimensional shock profile is implemented
\begin{align}
\label{eq:1Dadvection}
    u(x) = \frac{u_1 + u_2}{2} - \frac{u_1 - u_2}{2} \tanh\left( \frac{x}{L_{\mathrm{sh}}}\right), 
\end{align}
with upstream and downstream velocity, $u_1$ and $u_2 = u_1/q$, and shock width $L_{\mathrm{sh}}$. In the following, all units are normalized so that
\begin{equation}
    \tilde{x} = \frac{x}{x_0},\quad \tilde{u} = \frac{u}{u_0}, \quad \tilde{t} = \frac{t}{t_0}, \quad \tilde{\kappa} =  \frac{\kappa}{\kappa_0},\quad  \tilde{E} =  \frac{E}{E_0},
\end{equation} 
with $x_0/v_0 = t_0$ and $E_0$ being the energy injected at the shock. Thus, different physical scenarios can be modeled with the same simulation by adjusting the time scale. For the CRPropa3.2 simulations we used $x_0 = 1\,\mathrm{km}$ and $v_0 = 1\,\mathrm{m/s}$, so that $t_0 = 1000\,\mathrm{s}$ and $\kappa_0 = x_0 v_0 = 1000\,\mathrm{m^2/s}$.

\begin{figure}[tbp] 
    \centering 
    \includegraphics[width=.6\textwidth]{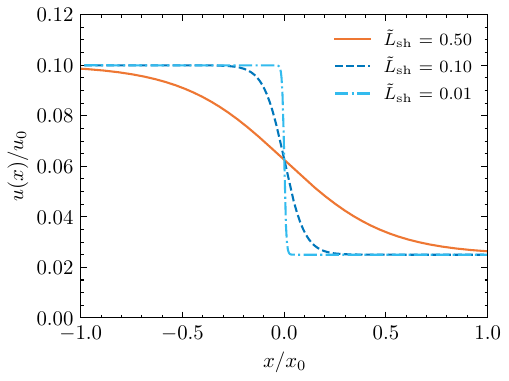}
    \caption{One-dimensional advection field with shock at $\tilde{x} = 0$ and a shock compression ratio $q = 4$. The smaller the shock width compared to the integration step length, the better the ideal shock solution is approximated.}
    \label{fig:1Dshockprofile}
\end{figure}

The velocity profile $u(x)$ is also used in other studies of linear and non-linear DSA \citep{KruellsAchterberg94, AchterbergSchure2010, WalterEA2022}. Figure \ref{fig:1Dshockprofile} illustrates $u(x)$ for a shock compression ratio $q = u_1/u_2 = 4$ and different shock widths. The narrower the shock width, the better the approximation of the ideal shock. However, from the numerical perspective the region where $\partial u / \partial x \neq 0$ must be large enough compared to the integration step length for candidates to experience acceleration. Otherwise energy gains may be underestimated (see Section \ref{sec:1DConstraints} for a detailed analysis).

    \subsection{Stationary Solution}
    \label{sec:1DStationary}

    Simulating with CRPropa 3.2, pseudo-particles are injected at $t=0$ at the shock and propagated until a maximum simulation time or until they leave the acceleration region downstream through free-escape boundaries. 

\begin{figure}[htbp] 
        \centering 
        \includegraphics[width=1\textwidth]{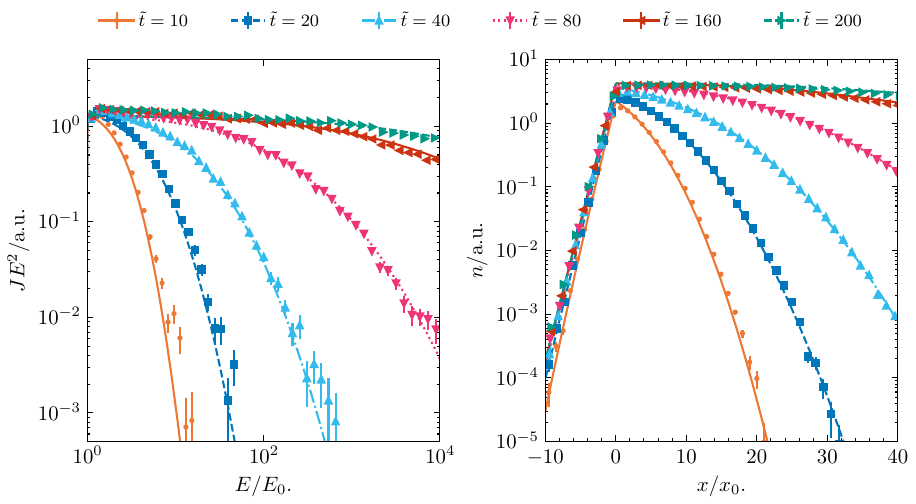}
        \caption{\label{fig:TimeEvolution_1D} Time evolution of spectrum at the shock, $\tilde{x} = [0, 2]$ (left) and number density, $n = \int \mathcal{N}\,\mathrm{d}p$, in the acceleration region, here defined by the free escape boundaries $L_{\pm}$} (right). Particles with energy $E_0$ are continuously injected at the shock from time $\tilde{t} = 0$. Upstream speed is $\tilde{u}_1 = 1$ with a compression of $q = 4$ at the shock. The diffusion coefficient is constant in space and energy, $\tilde{\kappa} = 1$. Free escape boundaries are at $\tilde{L}_{\pm} = \pm100$. Spectrum and number density resulting from simulation with CRPropa3.2 (dots) are compared to the solutions obtained by integrating the transport equation with VLUGR3 (lines).
    \end{figure}

    In order to obtain a stationary solution to compare the results with shock spectra predicted from theory, $s = 2 - 3q/(q-1)$ (e.g.\ \cite{Drury}), continuous injection upstream must be assumed. The inherent structure of CRPropa 3.2 does not allow for continuous injection during the simulation itself.
    To construct the stationary solution at the shock afterwards CRPropa's \texttt{TimeEvolutionObserver} module is used. Positions and energies of pseudo-particles are stored at times $t_{0} + n\Delta T$, which are called \textit{time snaps} and do not necessarily have to be equal to the simulation time step $\Delta t$. With the assumption of continuous injection, the solution at time $t$ is constructed by summing over all time snaps to approximate the time-integrated solution. This approach is valid given that the solution does not change too much during one time interval $\Delta T$. The time intervals $\Delta T$ can be chosen linearly or to increase with time, when the solution undergoes smaller changes. The resulting time evolution of the spectrum at the shock and number density integrated over momentum are shown in Fig.\ \ref{fig:TimeEvolution_1D} compared to the solutions obtained by the integration of the Fokker-Planck equation (\ref{eq:TransportEq}) using the same parameters. For the integration of the partial differential equations VLUGR3 \citep{VLUGI, VLUGII} is used. The spectra are weighted by energy so that the predicted slope lies horizontally in the plot. The resulting spectra and number densities are in good agreement. Differences may result from binning in energy and space for the CRPropa3.2 simulation and from resolution in energy and time for the VLUGR3 simulation. 
    
    The time evolution can be interpreted in the following ways: The shock is active from time $t = 0$ on and accelerated particles entering the shock region with energy $E_0$. The longer the shock is active the higher the maximum energy the particles can reach and the cut-off of the spectrum moves to higher and higher energies. 
    Having an infinite acceleration region, the stationary solution of a planar one-dimensional shock with compression $q = 4$ is a power-law with slope $s = -2$. With increasing simulation time, the approximation of the stationary solution gets better and the solution relaxes to the power-law for ever higher energies. In Fig.\ \ref{fig:TimeEvolution_1D} the stationary solution at $\tilde{t}=200$ is reached for energies $E < 10^3 E_0$. Theoretically there is no limit for the maximal energy, due to the infinitely large shock front every $n$-th crossing there will always be a nonzero number of particles to cross the shock an $n+1$-th time for $t \rightarrow \infty$. Considering a spherical geometry as discussed in Section \ref{sec:2DSpherical}, the maximal energy that can be reached depends on the diffusion coefficient and shock radius.

    In order to evaluate the approximation of the stationary solution the spectrum is fitted at $\tilde{t} = 500$ (not shown in Fig.\ \ref{fig:TimeEvolution_1D}) for energies below $10^4 E_0$. Up to that energy candidates are splitted by the \texttt{CandidateSplitting} module. The resulting spectral slope of $-2.054 \pm 0.003$ matches the expected value $s = -2$ nicely. 

    The time evolution of the number density shows how particles gradually migrate into the downstream region due to the advection field. At $\tilde{t}=200$ the stationary solution is reached close to the shock, $\tilde{x} < 30$, at later times also further into the downstream region. The free-escape boundary at $\tilde{x}=100$ does not influence the number density profiles at the shock. In principle, free escape boundaries are only needed if the acceleration region has a finite size. In Fig.\ \ref{fig:DiffusiveStepBoundary} we show how the resulting spectrum is affected by free-escape boundaries.
    Only a fraction of particles makes it into the upstream region $x < 0$ against the advection flow. Acceleration is efficient for a constant, low diffusion coefficient as assumed here, however, particles are unlikely to escape upstream. 
     
    The analytical solution by \citep{Toptyghin} is only for the special case of $v^2/\kappa = \mathrm{const.}$ across the shock, resulting in a drop of spatial diffusion by $1/16$ for a strong shock. Forman and Drury \citep{FormanDrury83} give an approximate solution for momentum dependent $\kappa$, which is exact if $v^2/\kappa = \mathrm{const.}$. Such spatial dependence of the diffusion coefficient is not considered here, since the resulting drift term which adds to the advective step in Eq.\ (\ref{eq:SDEI}) is currently not implemented in CRPropa3.2. For SDEs the strong additional drift term that is expected at the shock may also lead to numerical difficulties \citep{AchterbergSchure2010}. The integration of the transport equation using VLUGR3 is verified against existing analytical solutions in earlier works \citep{WalterEA2022} and by comparisons of the acceleration time scale \citep{Drury, FormanDrury83}. In Appendix \ref{sec:FormanApprox} the approximate time-dependent solution from \citep{FormanDrury83} is compared to the solution obtained with VLUGR3.

    \subsection{Constraints}
    \label{sec:1DConstraints}

    In order to reproduce the predicted spectrum the simulation set-up must fulfill constraints on the integration step length and shock width. For SDEs Achterberg and Kr\"ulls \citep{KruellsAchterberg94} and later Achterberg and Schure \citep{AchterbergSchure2010} present a thorough analysis of the choice of shock width and step length and the resulting spectral slope. The simulations resulting in Fig.\ \ref{fig:TimeEvolution_1D} take their findings into account. In the following we show that DSA simulations with CRPropa3.2 are subject to the same constraints. These constraints are not purely numerical but can also be motivated from physics.
    
    Considering a diffusion coefficient constant in space and energy, the advective step length, diffusive step length and shock width must fulfill the inequality
    \begin{equation}
    \label{eq:ineq}
        u \Delta t < L_{\mathrm{sh}} \lesssim \sqrt{\kappa \Delta t},
    \end{equation}
    to obtain the correct spectral index \citep{KruellsAchterberg94}. 
    
    The first inequality ensures that pseudo-particles experience the gradient of the advection field and therefore gain energy. Numerically, the time step must be chosen sufficiently small to resolve the shock region. On the other hand, the diffusive step length $\sqrt{\kappa \Delta t}$ --- a measure for the stochastic step --- must be larger than the shock width to increase the likelihood of pseudo-particles to cross the shock front multiple times in finite simulation time. Depending on the chosen shock width, diffusion coefficient and advection field it may not be possible to fulfill the inequality. In that case, acceleration can be under- or even overestimated. In Appendix \ref{sec:AdvectiveStep} the resulting spectral slopes for various shock widths and time steps are shown, with the physical parameters, advective speed and diffusion coefficient, being held constant.

    Constraints on the time step come from the numerical method, however, not all combinations of $u$, $L_{\mathrm{sh}}$ and $\kappa$ can be physically motivated. For instance, if diffusion is low compared to the advective speed there will be negligible acceleration.

    Obeying Eq.\ (\ref{eq:ineq}) alone is not sufficient to reproduce the ideal shock spectra. If the shock width is too large, pseudo-particles experience a smoothly changing advection field instead of a discrete shock. For diffusion independent on space and energy Kr\"ulls and Achterberg \cite{KruellsAchterberg94} calculated the expected spectral slopes for such smooth velocity gradients. In Fig.\ \ref{fig:TimeStepShockWidth} we show analogous to \cite{KruellsAchterberg94} that the expected spectral slopes for such smooth shock waves are met if the time step is sufficiently small. 
    
    Also, when a finite acceleration region is considered, diffusion must not only be large enough so that particles are able to diffusive back to the shock but also low enough so that they are contained in the acceleration region and do not escape too quickly. The effect on a finite acceleration region on the resulting spectral index is shown in Fig.\ \ref{fig:DiffusiveStepBoundary}. 

    In the following we show the effect of the chosen time step and shock width as well as the diffusion coefficient. We aim to reproduce the predicted spectral slope $s = 2 - 3q/(q-1)$ for compression ratios $q = 2$ and $q = 4$.

    \subsubsection*{Advective Step}
    
     Acceleration at shocks of different widths $\L_{\mathrm{sh}}$ is simulated with decreasing time steps $\Delta t$. The normalized diffusion coefficient is $\tilde{\kappa} = 1$ and the upstream velocity $\tilde{u}_1 = 1$. The shocks compression ratio is $q = 2$ leading to a spectral slope $s = -4$ in order to compare with the predictions from \cite{KruellsAchterberg94}. The slope is determined by linear fits in log-space up to the maximum energy of the \texttt{CandidateSplitting} module. Time is chosen such that the cut-off of the spectrum has minimal impact on the slope approximating the stationary solution. The results are shown in Fig.\ \ref{fig:AdvectiveDiffusiveStep}. 
     
     The expected spectral slopes depending on the shock widths are indicated by lines. With increasing resolution the simulated spectrum well approximates the predicted stationary solution. For sufficiently small shock widths, here $\tilde{L}_{\mathrm{sh}} = 0.008$, a spectral slope close to that of an ideal shock is obtained. 
     The larger the shock width, the less efficient the acceleration which leads to steeper spectra. Depending on the shock width, with decreasing time step, the simulation approaches the predicted stationary solution from steeper ($\tilde{L}_{\mathrm{sh}} = 0.128$) or flatter ($\tilde{L}_{\mathrm{sh}} = 0.008$) spectra. Thus, the compression ratio of the shock is over- or underestimated. 
     
     We show a more detailed figure with entangled information about the shock width and time step in the Appendix in an analogous way to \cite{KruellsAchterberg94}. 

    \subsubsection*{Diffusive Step}
    
    A similar analysis can be done for the dependence of the diffusion coefficient in relation to the shock width. The larger the diffusive step, the better becomes the approximation of an ideal shock. If the diffusive step however is too large, pseudo-particles may miss the shock region and acceleration is underestimated as a consequence. 
  
\begin{figure}[htbp] 
    \centering 
    \includegraphics[width=0.49\textwidth]{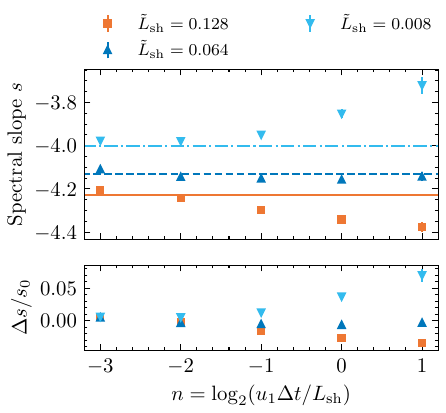}
    \includegraphics[width=0.49\textwidth]{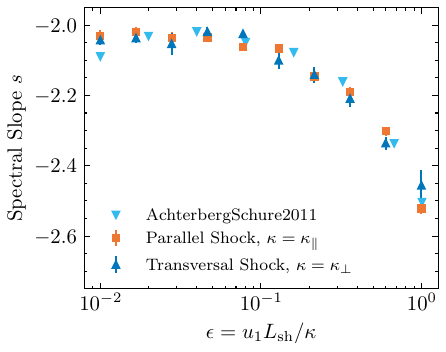}
    \caption{Left: Resulting spectral slope depending on the time step $\Delta t$ in relation to the shock width $L_{\mathrm{sh}}$. The upstream speed $\tilde{u}_1 = 1$. The compression ratio is $q = 2$, thus for an ideal shock we expect $s= -4$. With larger shock widths, the shock produces steeper spectra as calculated by \cite{KruellsAchterberg94}. Slopes of the predicted stationary solutions are indicated by the colored lines for the respective shock width. Right: Resulting spectral slope depending on the ratio of upstream velocity and shock width to diffusion coefficient. Data from \cite{AchterbergSchure2010} is shown for comparison. Here $\tilde{u}_1 = 1$, $q = 4$ and $\tilde{L}_{\mathrm{sh}} = 0.004$ with $\Delta \tilde{t} = 0.001$ is used while $\kappa$ is varied. The free escape boundaries are $\tilde{L}_{\pm} = 10$. With higher diffusion coefficient the spectral slope approaches the predicted value $s = -2$. We already show the results for a transversal shock wave and taking perpendicular diffusion into account, which is discussed in Section \ref{sec:2DTransversal}. }
    \label{fig:AdvectiveDiffusiveStep}
\end{figure}

    For a constant shock width of $\tilde{L}_{\mathrm{sh}} = 0.004$ the diffusion coefficient was varied over several simulation runs. The resulting spectral slope was fitted in the energy range $[E_0, 10^3 E_0]$ given by the \texttt{CandidateSplitting}. The results are shown in Fig.\ \ref{fig:AdvectiveDiffusiveStep} in comparison to the results from \cite{AchterbergSchure2010}. 
    With large diffusive step length the spectrum may also be affected by free-escape boundaries. For the simulations boundaries were chosen such that they do not influence the time evolution of the spectrum for the simulated diffusion coefficients. Likely, that is why we do not find the spectral slope to decrease again for $\epsilon < 0.04$. In the Appendix we show how the free-escape boundaries influence the spectrum at the shock depending on the diffusion coefficient. 

    We conclude that for the simulation of ideal shocks, the shock width must be small compared to the advective field and the time step should be chosen such that the advective step is at most as large as one fourth of the shock width. The diffusive step must be larger than the shock width, but not too large to still contain particles in the acceleration region. Those constraints may be difficult to fulfill when modeling a physical scenario and may require very small time steps close to the shock. 
    
    \subsection{Finite Lifetime}
    \label{sec:1DBurst}

    We model two scenarios that lead to a shock spectrum that is not stationary: Either the source of particles reaching the shock is only active for a short period of time or the shock itself has a finite lifetime. 
    Considering the GWTS, the CR flux that reaches the shock from the Galactic disk can be assumed to be stationary. The GWTS itself however cannot be maintained for an infinite time since too much mass would be advected out of the Galaxy. 
    
\begin{figure}[htbp] 
    \centering 
    \includegraphics[width=1\textwidth]{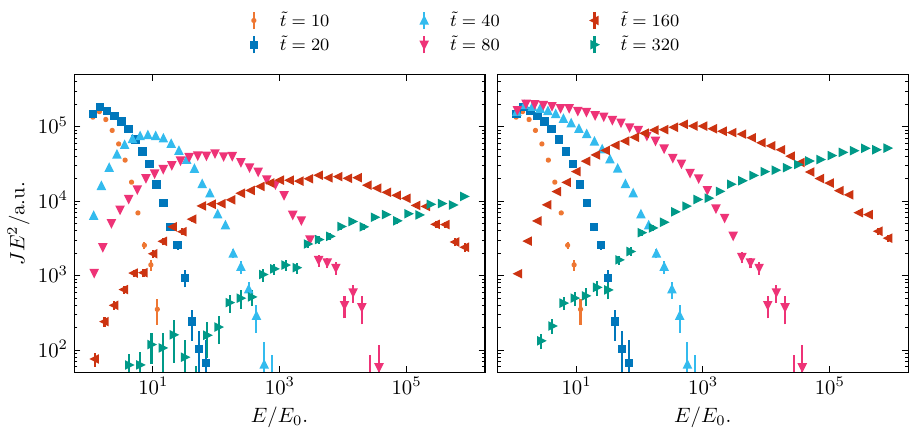}
    \caption{Candidates are injected at $\tilde{t}=0$ and simulated until they leave the acceleration region through free-escape boundaries $\tilde{L}_{\pm} = \pm 100$. A finite lifetime $\Delta \tilde{T}_{\mathrm{scr}} = 20$ (left) and $\Delta \tilde{T}_{\mathrm{scr}} = 80$ (right)  of the particle source is assumed. The spectrum at the shock $\tilde{x} = [0,2]$ is shown. }
    \label{fig:BurstInjection}
\end{figure}

    To model a burst-like particle source, candidates are injected at $t = 0$ and propagated until they leave the acceleration region through the free-escape boundaries $L_{\pm}$. The time evolution depending on the source duration can be derived similar to the stationary spectrum in Section \ref{sec:1DStationary}. Energy and position of candidates are stored in time intervals $\Delta T$ and the resulting spectrum is approximated by summing of those time snapshots from $[t-\Delta T_{\mathrm{src}}, t]$ with $\Delta T_{\mathrm{src}}$ being the source lifetime. For more details we point to \cite{MertenEA2018}. Figure \ref{fig:BurstInjection} shows the time dependent spectrum at the shock for two different source duration $\Delta T_{\mathrm{src}}$. 

\begin{figure}[htbp] 
    \centering 
    \includegraphics[width=1\textwidth]{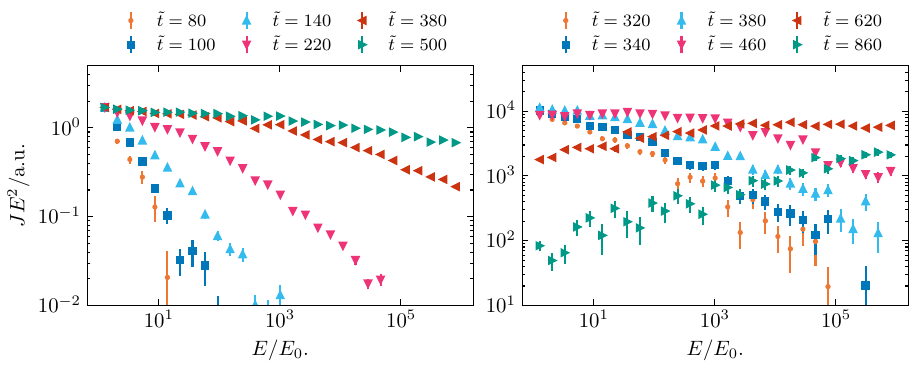}
    \caption{Left: Assuming continuous injection of candidates, the time evolution of the downstream spectrum at $\tilde{x} = 50$ is shown. Right: Candidates are injected at $\tilde{t}=0$ and simulated until they leave the acceleration region through free-escape boundaries $\tilde{L}_{\pm} = \pm 100$. A finite lifetime $\Delta \tilde{T}_{\mathrm{scr}} = 100$ of the shock is assumed. The spectrum of particles escaping through the downstream boundary is shown.}
    \label{fig:FiniteShock}
\end{figure}

    Considering a finite lifetime of the shock, we already know the time dependent spectrum at the shock from Section \ref{sec:1D}. In order to assess the contribution to the overall cosmic-ray spectrum, not only the spectrum at the shock, but also the spectrum of particles escaping downstream, or even upstream is of interest. Fig.\ \ref{fig:FiniteShock} (left) shows the time evolution of the downstream spectrum, $\tilde{x} = 50$, assuming continuous injection at the shock. Compared to Fig.\ \ref{fig:TimeEvolution_1D} the spectrum evolves later, since particles need time to get from the shock to the downstream position. Considering now a free-escape boundary at $\tilde{L}_{+} = 100$, Fig.\ \ref{fig:FiniteShock} (right) shows the time-dependent spectrum of particles escaping with a lifetimes of the shock $\Delta \tilde{T}_{\mathrm{sh}} = 100$. For the first time steps, the spectrum still evolves up to the quasi-stationary solution ($\tilde{t} = 450$). Particles that stay long in the acceleration region, defined by the free escape boundaries $L_{\pm}$, reach higher energies than particles that escape quickly. Thus, at late times ($\tilde{t} = 620, 860$), the spectrum gets flatter since low energy particles already escaped. 
    
    With a constant, low diffusion coefficient $\tilde{\kappa} = 1$ acceleration is efficient but particles do not escape upstream. In the following Section we introduce energy-dependent diffusion: With increasing energy, diffusion is higher and particles are more likely to escape upstream.

    \subsection{Energy-dependent Diffusion}
    \label{sec:1DEnergy}

     So far, a diffusion coefficient constant in energy was considered. A more realistic description of diffusive motion takes energy-dependence into account.
     Assuming that the diffusion coefficient is proportional to $(E/E_0)^{\alpha}$ and $\alpha > 0$, the diffusive step becomes larger with increasing energy. At first glance it is then easy to satisfy the inequality in Eq.\ (\ref{eq:ineq}). However, the analysis in Section \ref{sec:1DConstraints} revealed that the diffusive step should not become too large and free-escape boundaries should be dropped or must be set far away from the shock, otherwise high-energy particles escape too quickly. 
     
     In order to keep the diffusive time step within a reasonable range, an energy-dependent adaptive time step is implemented in the \texttt{DiffusionSDE} module. Within a range of specified time steps $[\mathrm{d}t_{\mathrm{min}}, \mathrm{d}t_{\mathrm{max}}]$, it is chosen such that the advective step is smaller than $1/4\,L_{\mathrm{sh}}$ and the diffusive time step smaller than $100\,L_{\mathrm{sh}}$. This also leads to better performance in case of the energy independent diffusion coefficient ($\alpha = 0$), since the time step in the downstream region can be chosen higher than in the upstream region by a factor of the shock compression ratio $q$. 

\begin{figure}[htbp] 
    \centering 
        \includegraphics[width=1\textwidth]{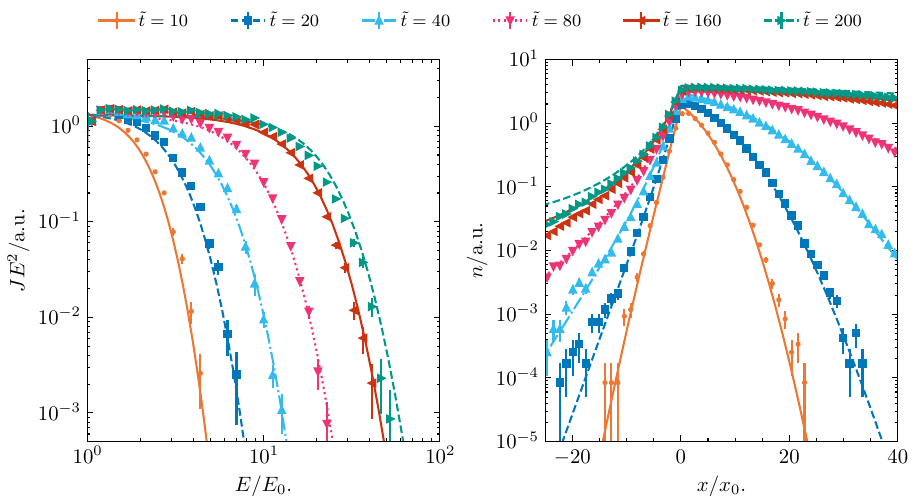}
        \caption{Time evolution of the spectrum (right) at the shock and the number density (left) close to the shock with energy-dependent diffusion coefficient $\kappa = \kappa_{E_0} \left(E/E_0\right)$ compared to the results obtained by integrating the transport equation using VLUGR3. }
    \label{fig:EnergyDependenceTimeEvolution}
    \end{figure}

    Figure \ref{fig:EnergyDependenceTimeEvolution} shows the time evolution of the spectrum at the shock and number density in the acceleration region. Again the compression ratio is $q = 4$ and the predicted spectral slope $s = -2$. The diffusion coefficient is $\kappa(E) = \kappa_{E_0} \left(E/E_0\right)^{\alpha}$ with $\alpha = 1$ and $\tilde{\kappa}_{E_0} = 1$. The results obtain with the diffusion approach applied in CRPropa3.2 are compared to the simulations using VLUGR3. The same times as in Fig.\ \ref{fig:TimeEvolution_1D} are used.

    With $\alpha > 0$ the diffusion coefficient $\kappa(E)$ is equal or greater than in the energy-independent case shown in Section \ref{sec:1DStationary}. With that the acceleration time scale gets larger. The cut-off of the energy spectrum at the shock is at lower energies compared to Fig.\ \ref{fig:TimeEvolution_1D}. 

    With higher diffusion at high energies, more particles make it in the upstream region against the advection flow. Since high energies are only reached at later times, the upstream number density also evolves over time in contrast to Fig.\ \ref{fig:TimeEvolution_1D} for energy-independent diffusion.
    Thus, the likelihood of escaping upstream gets higher with energy, thus acceleration time, and increasing $\alpha$. 

    The cut-off energy for a planar shock is still only determined by the free-escape boundaries and acceleration time. Here no free escape boundaries are used, so that the finite acceleration region does not influence the spectrum at the shock. In physical scenarios, high energy particles are likely to escape the acceleration region. 
    
\begin{figure}[htbp] 
    \centering 
        \includegraphics[width=1\textwidth]{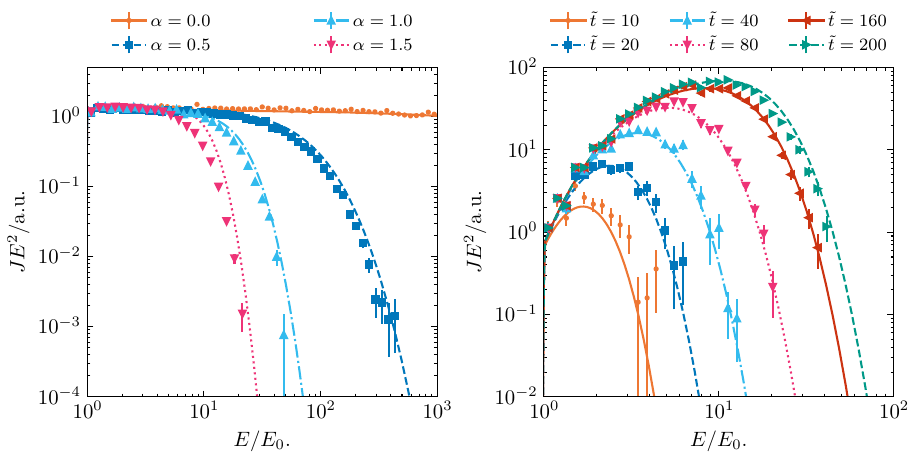}
        \caption{Left: Spectrum at the shock at time $\tilde{t} = 200$ with energy-dependent diffusion $\kappa = \kappa_0 \left(E/E_0\right)^{\alpha}$ and different values of $\alpha$. Energy spectrum obtained with CRPropa3.2 (dots) is compared against those resulting from VLUGR3 (lines). Right: Upstream spectrum, $\tilde{x} = [-6, -5]$, with energy-dependent diffusion ($\alpha = 1$). For energy-dependent diffusion the upstream spectrum differs from the downstream spectrum. Again CRPropa3.2 results are compared to VLUGR3.}
    \label{fig:EnergyDependenceUpstream}
    \end{figure}

    In Fig.\ \ref{fig:EnergyDependenceUpstream}a the spectra at the shock for different values of $\alpha$ are compared at time $\tilde{t} = 200$. The stronger the energy dependence, the higher the diffusion at high energies and the longer it takes to accelerate particles to the same energy.
    
    In case of energy-dependent diffusion, the upstream spectrum differs from the downstream spectrum. In Fig.\ \ref{fig:EnergyDependenceUpstream}b we show the time evolution of the upstream spectrum at $\tilde{x} = -5$ for $\alpha = 1$. 

    \subsection{Acceleration Time Scale}
    \label{sec:AccTimeScale}

    The average time to accelerate up to momentum $p$ is compared to the expression for the mean acceleration time \citep{Drury} 
    \begin{align}
    \label{eq:accTime}
        \bar{t} = \frac{3}{u_1 - u_2} \int_{p_0}^{p} \left( \frac{\kappa_1}{u_1} + \frac{\kappa_2}{u_2} \right)\, \frac{\mathrm{d}p'}{p}
    \end{align}
    for energy-independent and energy-dependent diffusion coefficients $\kappa_1 = \kappa_2 = \kappa$.
    
\begin{figure}[htbp] 
    \centering 
        \includegraphics[width=1\textwidth]{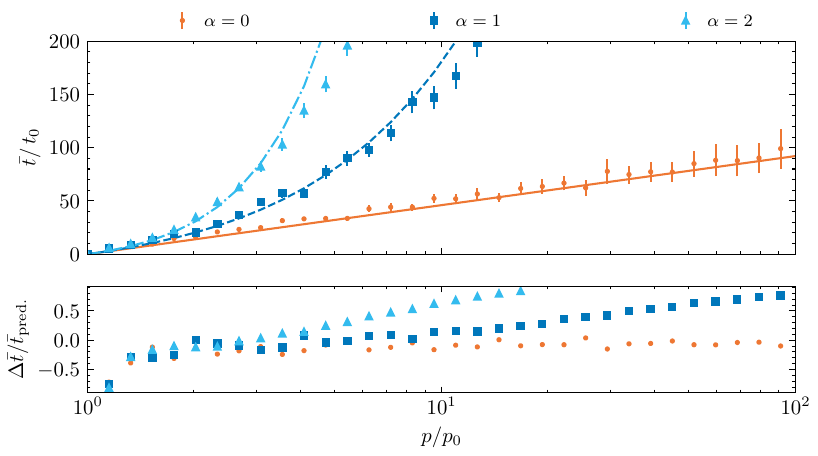}
        \caption{Top: Mean acceleration time up to momentum $p$ for diffusion coefficients with energy-dependence $\alpha = 0$ (solid line), $\alpha = 1$ (dashed line) and $\alpha = 2$ (dash-dotted line). Results obtained from CRPropa simulations are plotted for comparison. Error bars show the error of the mean time. To calculate the mean acceleration time up to $\bar{t} = 200\,t_0$, simulations were run until $T = 500\,t_0$. Bottom: Relative deviation of the simulated acceleration time to the prediction. With higher momentum the deviation gets larger due to finite simulation time, especially for energy-dependent diffusion.}
    \label{fig:accTime}
    \end{figure}

   To estimate the mean acceleration time, candidates are injected at $\tilde{t} = 0$ at the shock and accelerated up to a simulation time $T$ and energy and position are stored in time steps $\Delta T$. Candidate splitting is used to increase the statistics at high energies. At the shock, $\tilde{x} = [0,1]$, for bins in momentum $p_i = E_i/c^2$, the average times $\bar{t}(p_i)$ are calculated. Figure \ref{fig:accTime} compares the calculated mean acceleration time from simulations with energy-dependence $\alpha = 0, 1, 2$ of the diffusion coefficient with the expectation derived from Eq.\ (\ref{eq:accTime}). 

    In general, the mean acceleration time increases with momentum. With momentum dependent diffusion coefficient, acceleration gets slowed down over time, leading to different dependencies of the mean acceleration time on momentum
    \begin{align}
    \label{eq:MeanAccTime}
        \bar{t}(p) \propto \begin{cases} \tau_{\mathrm{acc}} \, \mathrm{ln}\left( \frac{p}{p_0}\right), &  \alpha = 0 \\
         \tau_{\mathrm{acc}} \left( \frac{p}{p_0} - 1\right), & \alpha = 1 \\
        \tau_{\mathrm{acc}} \left[ \left(\frac{p}{p_0}\right)^2 - 1\right], & \alpha = 2,
        \end{cases}
    \end{align}
    with $\tau_{\mathrm{acc}} = 3/(u_1 - u_2) (\kappa_1/u_1 + \kappa_2/u_2)$. In general, the mean acceleration time fits the expected behavior. In Appendix \ref{sec:FormanApprox} we also compare the time-dependent shape of the spectrum obtained with VLUGR3 to the approximation derived by \citep{FormanDrury83}.

    \subsection{Injection of Pre-accelerated Spectra}
    \label{sec:1DPreAcc}

    So far mono-energetic pseudo-particles of energy $E_0$ have been injected at the shock. In a physical scenario of particle acceleration at the GWTS, particles are assumed to arrive with an energy distribution of a power law \citep{ThoudamEA2016}. Depending on the acceleration process a range of spectral slopes is possible. 

\begin{figure}[htbp] 
    \centering 
        \includegraphics[width=.49\textwidth]{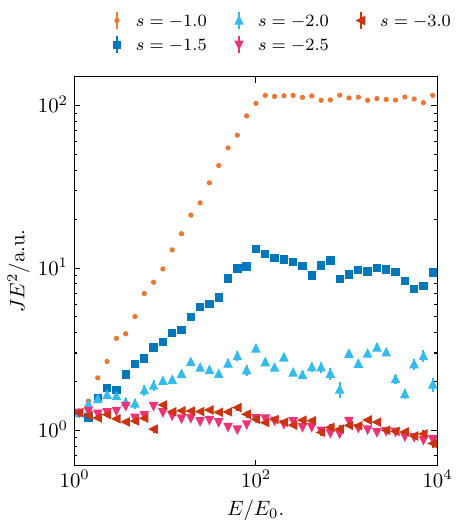}
        \caption{ Spectra at $\tilde{t} = 400$ at the shock for different injected spectra from $10^{-1} E_0$ up to $10^2 E_0$. For $s < -2$ the spectrum converges to the shock spectrum. For $s > -2$ the spectrum breaks at $10^2 E_0$.  }
    \label{fig:InjectSpectra}
\end{figure}

    We injected different spectra at the shock and investigated the impact on the resulting re-accelerated spectra. 
    According to \cite{Drury} steeper spectra than the one produced by the shock converge to the shock spectrum. When, on the other hand, the shock produces a steeper spectrum than the one continuously injected, the injected spectrum does not change. This results in broken power-law spectra when the injected spectrum has a finite cut-off energy as shown in Fig.\ \ref{fig:InjectSpectra}. 

    \begin{figure}[htbp] 
    \centering 
        \includegraphics[width=1\textwidth]{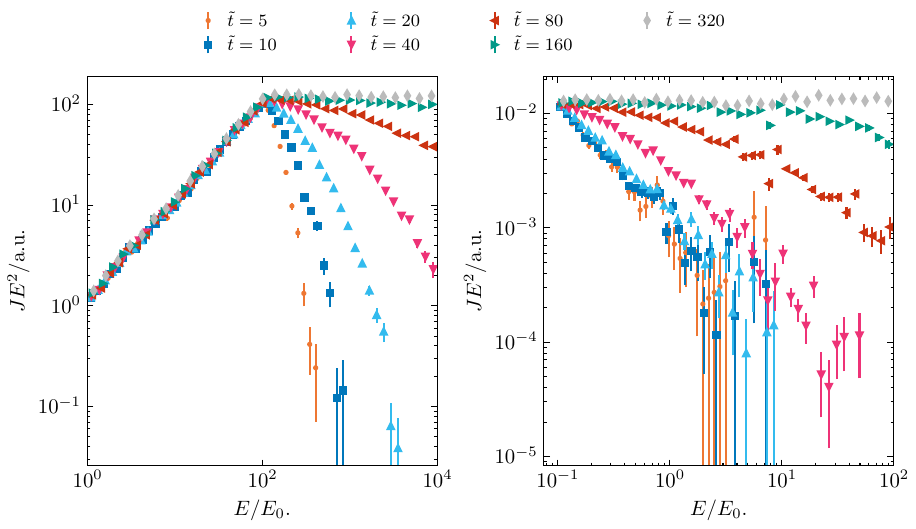}

        \caption{Time evolution of the spectrum at the shock $\tilde{x} = [0,2]$ with injected spectra of slope $s = -1$ (left) and $s = -3$ (right). The injected spectra have a hard cut-off at $10^2 E_0$. For $s = -1$ up to $10^2 E_0$ the spectral slope remains the same, for higher energies the spectrum evolves to the one produced by the shock, $s = -2$. For $s = -3$ the spectrum converges to the shock spectrum for all energies.   
        }
    \label{fig:InjectedSpectra-TimeEvo}
\end{figure}

    In Fig.\ \ref{fig:InjectedSpectra-TimeEvo}a a flatter spectrum than produced by the shock is continuously injected up to the maximum energy $10^2\,E_0$. Up to that energy, the injected spectrum prevails. At the shock, particles are accelerated independent on their energy and the time evolution of the spectrum at the shock becomes visible for $E > 10^2\,E_0$. In Fig.\ \ref{fig:InjectedSpectra-TimeEvo}b a steeper spectrum than produced by the shock is continuously injected up to the maximum energy $10^2\,E_0$. The spectrum converges to the one produced by the shock over time, since all particles independent on their energy experience acceleration. Thus, at the shock particles are re-accelerated to higher energies.
    
     In case of flatter injected spectra, broken power-law spectra may emerge. In case of steeper spectra, the spectrum converges to the shock spectrum. If the injected spectrum is the same as produced by the shock it remains the same. This is summarized in Fig.\ \ref{fig:InjectSpectra}: The stationary spectrum up to $10^4 E_0$ at the shock with different injected spectra from $10^{-1} E_0 - 10^2 E_0$ is shown. It is more likely that the injected spectrum has an exponential cut-off. In that case, the cut-off behaves as a steep spectrum and will simply accelerated up to the shock spectrum. 
    
\section{Anisotropic Diffusion, Oblique Shocks and Spherical Symmetry}
\label{sec:2D}

In the following, we extend our analysis to anisotropic diffusion, oblique planar shocks and spherical symmetry. The diffusion coefficient is held constant in space and energy. Note that since the diffusion coefficient is defined in the local magnetic field coordinates, diffusion changes with the direction of the magnetic field even when $\kappa$ is held constant. In Section \ref{sec:2DSpherical} a spherical shock profile with a spiral magnetic field is modeled. 

    \subsection{Perpendicular Shock Wave}
    \label{sec:2DTransversal}

    We consider a perpendicular shock wave with the magnetic field perpendicular to the advection flow. If only parallel diffusion is considered, particles would not be able to cross the shock and, therefore, do not experience acceleration. The advective field is still given by Eq.\ (\ref{eq:1Dadvection}), the magnetic field now points in y-direction, $\vec{B} = B_0\vec{e}_{\mathrm{y}}$. 
    
    The perpendicular diffusion coefficient now simply takes the role of parallel diffusion. We can find the same dependency between perpendicular diffusion $\kappa_{\perp} = \epsilon \kappa_{\parallel}$ and shock width as shown in Fig.\ \ref{fig:AdvectiveDiffusiveStep} in Section \ref{sec:1DConstraints}. The parallel part of the diffusion tensor does thus not influence the acceleration process.
    
    \subsection{Oblique Shock Wave}
    \label{sec:2DOblique}
    
    First, only the advection field is considered to be oblique and the magnetic field remains parallel to the shock normal. Only the parallel component to the shock normal of the advection field, $u_{\mathrm{x}}$, is shocked, so that it is now given by
    \begin{align}
    \label{eq:2Dadvection}
    u_{\mathrm{x}} &= \frac{u_{\mathrm{x},1} + u_{\mathrm{x},2}}{2} -  \frac{u_{\mathrm{x},1} - u_{\mathrm{x},2}}{2} \tanh\left( \frac{x}{L_{\mathrm{sh}}}\right), \\
    u_{\mathrm{y}} &= \mathrm{const.} 
    \end{align}
    The $y$-component is varied from $\tilde{u}_{\mathrm{y}} = 0$ (one-dimensional case) to $\tilde{u}_{\mathrm{y}} = 5 \tilde{u}_{\mathrm{x,1}}$, while $\tilde{u}_{\mathrm{x,1}} = 1$. The angle $\theta$ between shock normal and advection flow varies from $\theta_1 = 0$ to $\theta_1 \approx 0.44 \pi$. Here, the electric field is neglected, which results from the advection field and magnetic field being at an angle. In the SDE approach, the electric field would contribute to a drift in momentum which adds to Eq.\ (\ref{eq:SDEII}). Such drifts are currently not implemented in CRPropa3.2.
    
    From theory, the spectrum still depends only on the shock compression ratio $q = u_1\mathrm{cos}\theta_1/u_2\mathrm{cos}\theta_2$ until the shock vanishes for $\theta \rightarrow \pi/2$. Considering the constraints on the time step and shock width, particles now experience varying "effective" shock widths when they cross the shock with different angles to the shock normal. Thus, it becomes difficult to obtain the expected shock spectra for large angle $\theta$. The effective shock width can be written as $L_{\mathrm{sh, eff}} = L_{\mathrm{sh}}/\mathrm{cos}\theta$.
    With a shock width of $\tilde{L}_{\mathrm{sh}} = 0.001$ and adaptive time step $\Delta t
    \in \left[L_{\mathrm{sh}}/4, L_{\mathrm{sh}}/16\right]$ the predicted spectral slope $s = -2$ is obtained. The time evolution of the spectrum at the shock is independent on the angle between the shock and advection field and the amount of perpendicular diffusion to the magnetic field. Candidates are now injected upstream $\tilde{x}_{\mathrm{src}} = -1$, so that they are not trapped inside the shock for large angles of $\theta$.

\begin{figure}[htbp] 
    \centering 
    \includegraphics[width = 1\textwidth]{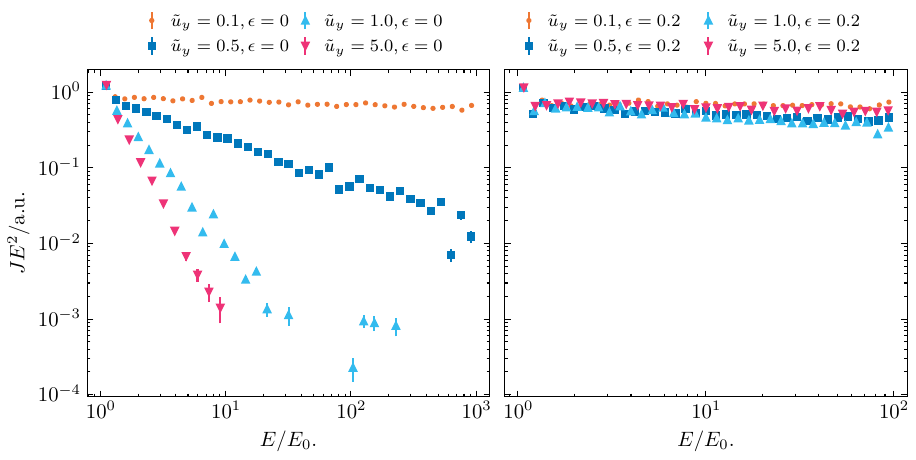}
        \caption{Resulting spectra at the shock for different angles between advection field and shock normal at $\tilde{t} = 400$. The magnetic field is considered parallel to the advection field. With parallel diffusion only (left) and increasing angle between magnetic field and shock normal, efficient acceleration becomes difficult and the predicted spectral slope $s = -2$ is only obtained for small angles. Introducing a small amount of perpendicular diffusion (right) with $\tilde{\kappa}_{\perp} = 0.2$ leads to the predicted spectral slope. }
    \label{fig:ObliqueSpectra}
    \end{figure}

     Now the magnetic field is considered to be parallel to the advection field. Both the advection field and magnetic field break at the shock. Here, curvature drifts due to the bent magnetic field lines at the shock are neglected. Currently drift terms are not included in the implementation of the diffusion coefficient in CRPropa3.2. Still, we expect to see acceleration at the shock up to the predicted spectrum. Figure \ref{fig:ObliqueSpectra} shows that with parallel diffusion only for large angles between magnetic field and shock normal, the spectral slope at the shock is steeper than predicted as acceleration is less efficient. However, introducing diffusion perpendicular to the magnetic field, particles are again able to cross the shock even if the magnetic field is almost perpendicular to the shock normal. A spectral slope $s = -2$ is obtained even for almost perpendicular shocks. 
    
    \subsection{Spherical symmetry}
    \label{sec:2DSpherical}

    We construct a simplified model of the GWTS, namely a spherical symmetric shock with spiral magnetic field that takes the findings of the previous sections into account. 
    With large shock radii astrophysical shocks can in principle be approximated by planar shocks with all implications described previously. However, for sufficiently small shock radii e.g.\ \cite{Drury} showed, that the spectral slope in spherical geometry is influenced by the energy- and spatial dependence of the diffusion coefficient. 
    In contrast to the one-dimensional planar shock, for spherical shocks a limit for the maximum energy is expected, since the shock front is not infinitely large. 

\begin{figure}[htbp] 
    \centering 
    \includegraphics[width = 0.6\textwidth]{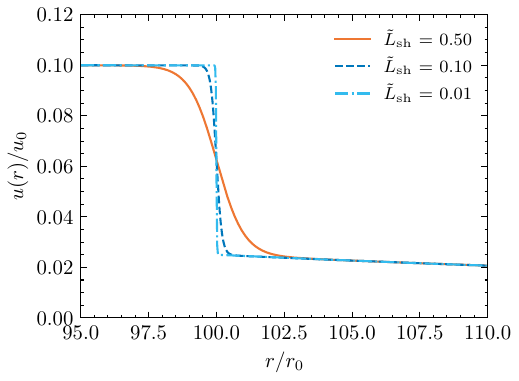}
        \caption{Radial shock profile. The advective field is constant upstream $u(r) = u_1$, drops by $1/q$ at the shock and decreases downstream with $r^{-2}$ for $r > R_{\mathrm{sh}}$. }
    \label{fig:radialshock}
    \end{figure}

    In the following, the advection field is assumed to be constant upstream, $r < r_{\mathrm{sh}}$, and to decrease with $r^{-2}$ downstream, $r > r_{\mathrm{sh}}$, modeled by
    \begin{align}
    \label{eq:RadialAdvection}
    \vec{u}(r) = u_1 \left\lbrace 1 + \left[ \left( \frac{R_{\mathrm{sh}}}{2r} \right)^2 \frac{1}{1 + \mathrm{e}^{-\frac{r-R_{\mathrm{sh}}}{L_{\mathrm{sh}}}}} \right] \right\rbrace \vec{e}_{\mathrm{r}}.
    \end{align}
    Analogous to Eq.\ (\ref{eq:1Dadvection}), a finite shock width $L_{\mathrm{sh}}$ is considered.
    Similar profiles were also used in other studies of spherical DSA \citep{FlorinskiJokipii2003} and applied to the GWTS \citep{JokipiiMorfill87, MertenEA2018}. After acceleration, the wind expands with constant speed $u(r)$. Particles experience adiabatic cooling upstream, $r < R_{\mathrm{sh}}$. Downstream, $r > R_{\mathrm{sh}}$ the wind is subsonic and decreases with $u(r) \propto r^{-2}$ without cooling. The advection field is shown in Fig.\ \ref{fig:radialshock} for different shock widths. 

\begin{figure}[htbp] 
    \centering 
    \includegraphics[width = 0.32\textwidth]{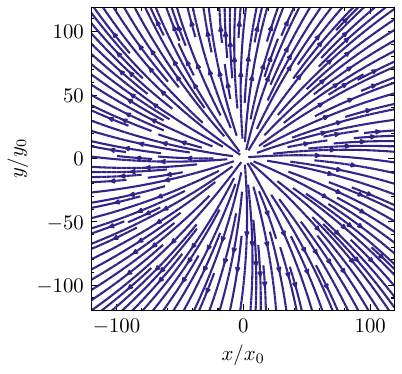}
    \includegraphics[width = 0.32\textwidth]{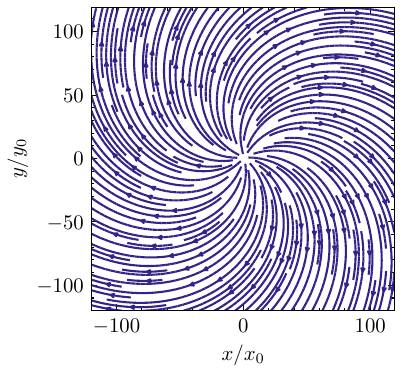}
    \includegraphics[width = 0.32\textwidth]{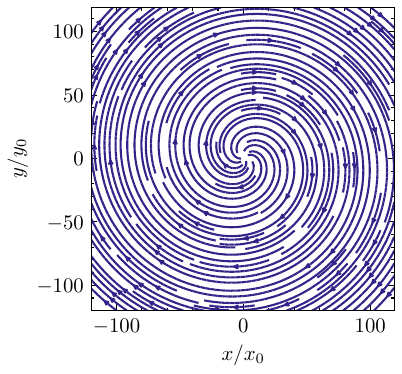}
    \caption{Face on view of the Archimedean Spiral magnetic field for increasing values of $\Omega = [0.1, 1.0, 10]$ from left to right. For small angular velocities the magnetic field is parallel to the shock normal, for high angular velocity magnetic field lines become perpendicular. }
    \label{fig:ArchimedeanSpiral}
    \end{figure}

    For the magnetic field, analogous to \cite{MertenEA2018}, an Archimedean spiral is considered. It is given by
    \begin{align}
    \vec{B} = \pm B_0 \left[ \left( \frac{r_0}{r} \right)^2 \vec{e}_{\mathrm{r}} - \frac{\Omega r_0^2 \sin{\theta}}{r v_{\mathrm{w}}} \vec{e}_{\phi} \right]
    \end{align}
    with the constant wind velocity $v_{\mathrm{w}}$ and $\Omega r_0 \sin{\theta}$ being the rotational velocity at a reference radius $r_0$ at latitude $\theta$. The magnetic field lines for different values of $\Omega$ are illustrated in Fig.\  \ref{fig:ArchimedeanSpiral}. For $\Omega = 0$ the magnetic field is parallel to the shock normal and the advection field. For $\Omega \neq 0$ the magnetic field is not parallel to the wind profile and the break of the magnetic field at the shock is neglected. For high values of $\Omega$ the magnetic field is almost perpendicular to the shock normal and resembles the transversal shock wave investigated in Section \ref{sec:2DTransversal}. 
    
\begin{figure}[htbp] 
        \centering 
        \includegraphics[width=1\textwidth]{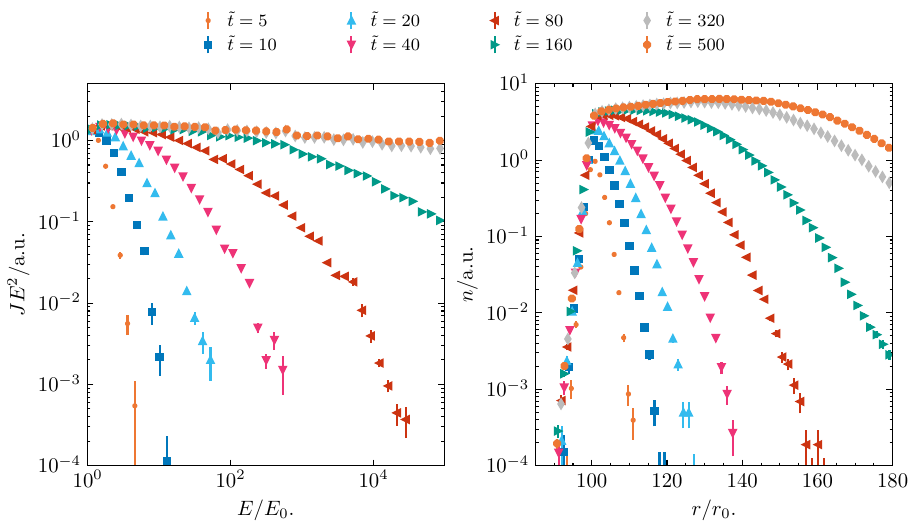}
        \caption{\label{fig:SphericalSolution} Time evolution of the number density (right) and spectrum at the shock, $\tilde{r} = [\tilde{r}_{\mathrm{sh}}, \tilde{r}_{\mathrm{sh}} + 2]$ (left). Free escape boundary is downstream $\tilde{R}_{+} = 200$. Both parallel and perpendicular diffusion are considered $\tilde{\kappa}_{\parallel} = 1$, $\tilde{\kappa}_{\perp} = 0.8$ and the magnetic field is almost perpendicular to the shock normal, $\Omega = 10$. }
    \end{figure}

    In Fig.\ \ref{fig:SphericalSolution} spectrum and number density are shown for $\Omega = 10$. Here, perpendicular diffusion was also taken into account to achieve sufficient acceleration. 
    In both simulations the stationary shock spectrum matches the predicted spectral slope $s = -2$ for a strong shock. Due to the decelerating flow for $r > R_{\mathrm{sh}}$ the number density profile differs from the planar scenario.

\begin{figure}[htbp] 
        \centering 
        \includegraphics[width=1\textwidth]{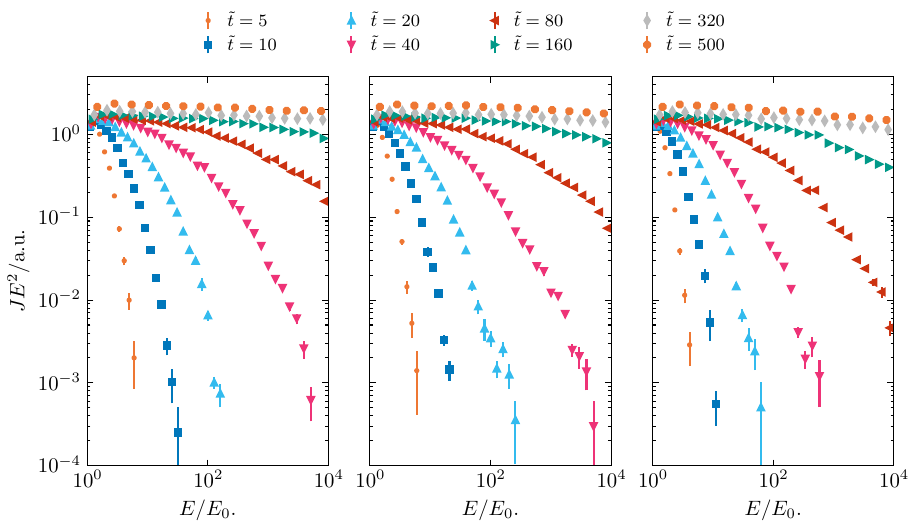}
        \caption{\label{fig:SphericalVariousEps} Time evolution of the spectrum at the shock, $\tilde{r} = [\tilde{r}_{\mathrm{sh}}, \tilde{r}_{\mathrm{sh}} + 2]$ for different values of perpendicular diffusion, $\epsilon = [0, 0.2, 0.8]$ from left to right. Free escape boundary is downstream $\tilde{R}_{+} = 200$. The lower the perpendicular diffusion, the slower the time evolution at the shock. }
    \end{figure}

    Figure \ref{fig:SphericalVariousEps} shows the time evolution of the spectrum at the shock and the number density in the acceleration region for $\Omega = 1$. Different values for the perpendicular diffusion are considered. Introducing diffusive motion perpendicular to the magnetic field, leads to higher diffusion in total. The time evolution of the spectrum at the shock is slower with higher perpendicular diffusion.

\section{Summary and Outlook}
\label{sec:summary}

In this paper we present a thorough analysis of modeling DSA with the \texttt{DiffusionSDE} module of CRPropa as a basis for future studies of the GWTS. 
CRPropa3.2 provides solvers for both transport regimes, and thus offers the opportunity to model acceleration at the GWTS and propagation back to the Galaxy in a coherent framework. 
Taking into account the constraints discussed in Section \ref{sec:1DConstraints}, time-dependent acceleration at shocks in three dimensions can be simulated with CRPropa3.2. Arbitrary magnetic field and advection field configurations can be used. Diffusion can be constant or energy-dependent as well as anisotropic. The new \texttt{CandidateSplitting} module speeds up the simulation significantly. 

We validated the resulting time-dependent spectra at a one-dimensional planar shock against those obtained from integrating the Fokker-Planck equation with the finite difference solver VLUGR3 and presented first applications introducing energy-dependent and anisotropic diffusion as well as spherical shock geometry.

With the \texttt{Acceleration} module, CRPropa3.2 also offers the possibility to model DSA in the test-particle regime. This was first introduced for second-order Fermi acceleration \citep{WinchenBuitink2018} and later extended to DSA \citep{CRPropa3.2}. Each time a particle crosses the shock front, the module calculates the energy gain performing Lorentz transformations from the particle frame to the rest frame of the scatter centers, the upstream or downstream background flow. However, this approach is less flexible than the simulation based on the transport equation or the corresponding SDEs. 

In Section \ref{sec:1D} we show that considering a one-dimensional planar parallel shock, simulations lead to the expected spectral slope at the shock. We validated the correct time evolution of the spectrum at the shock by comparing with simulations based on the integration of the transport equation (see \citep{WalterEA2022}). 
In Section \ref{sec:1DConstraints} we summarize constraints on the shock width, diffusion and time step first described by \cite{KruellsAchterberg94} and \cite{AchterbergSchure2010}. We find the same dependence between spectral slope, shock width and advective step, as well as diffusive step. Likewise we found that depending on the shock width for too small time steps, the compression ratio of the shock is over- or underestimated (see Fig.\ \ref{fig:AdvectiveDiffusiveStep} and Fig.\ \ref{fig:DetailedAdvectiveStep}). Presumably, this is due to the pseudo-particles only meeting the shock region by chance: The smaller the shock width the stronger the velocity gradient they encounter. This leads to harder spectra at least for those pseudo-particles experiencing acceleration.
Section \ref{sec:1DConstraints} explains in detail how to set up DSA simulations using SDEs in general and CRPropa3.2 specifically. 

Based on the findings in Section \ref{sec:1DConstraints} for DSA an alternative adaptive time step was implemented in the \texttt{DiffusionSDE} module. This ensures that within a user specified range the largest possible time step based on the inequalities (Eq.\ (\ref{eq:ineq})) for advective and diffusive step is used. Another extension to CRPropa3.2 is the \texttt{CandidateSplitting} module. High energies are only reached by a small fraction of candidates. For better statistics at high energies, candidates crossing energy boundaries are split into $n$ copies depending on the expected spectral slope. This way of importance sampling significantly reduces computation time. 

We approached more physical scenarios by considering energy-dependent diffusion, pre-accelerated spectra and shocks with finite lifetime. Depending on the life-time of the shock, the spectrum seen by a downstream observer differs from the stationary spectrum. Energy or spatial dependence of the diffusion coefficient would further modulate the downstream spectrum over time.
The effects on the shock spectrum discussed in sections \ref{sec:1DBurst} to \ref{sec:1DPreAcc} were deliberately considered separately and are the basis for future combined studies, potentially adding spatial dependent diffusion.

In Section \ref{sec:2D} we took anisotropic diffusion into account and showed that planar oblique shocks produce decent results, when diffusion parallel to the shock normal is high enough. 
A prototype for simulating the GWTS was presented in Section \ref{sec:2DSpherical}: A radial shock profile and a Archimedean spiral magnetic field. The time-dependent spectrum obtained at the GWTS as well as at the upstream and downstream escape boundaries can be used in further studies, investigating the portion of particles that is able to propagate back to the Galaxy (as done for a stationary shock spectrum by \cite{MertenEA2018}). Only high-energy CRs are able to propagate against the outflowing wind, so that energy-dependent diffusion must be considered. We already saw in Section \ref{sec:1DEnergy} that energy-dependent diffusion leads to a higher fraction of particles being able to diffusive back into the upstream region against the wind. 

For particles leaving the acceleration region the diffusion approximation might not be valid, depending on the escape time, magnetic field of the IGM and the particles' energy. At this point we stress again the advantage of CRPropa3.2: Follow up simulations can be done using both diffusive and ballistic propagation depending on the candidates energy within the same framework. 

Not only energy-dependent diffusion but also a spatial dependence might impact the acceleration process as well as the cosmic-ray transport in the Galactic magnetic field. A spatially varying diffusion tensor, however, induces an additional drift term to Eq.\ (\ref{eq:SDEI}) which is not yet implemented in CRPropa3.2. Also, this term adds to the left-hand side of the inequality in Eq.\ (\ref{eq:ineq}) discussed in Section \ref{sec:1DConstraints}. This makes it more difficult to choose a valid time step for a given shock width. Achterberg and Schure \cite{AchterbergSchure2010} even found that a second-order scheme might be necessary for handling the strong drift terms that would occur at the shock when the diffusion coefficient drops over the shock. 

The GWTS is most likely not a perfect sphere but like the termination shock of the heliosphere a spheroid. Also, it does not necessarily enclose the complete Galaxy. With a more variable wind profile CRs may even propagate back to the Galaxy more easily. A more realistic wind profile presumably changes the time evolution of the spectrum produced by the shock itself. When a finite shock lifetime is considered this may have an impact on the contribution to the shock spectrum at Earth.

With the radial wind profile and Archimedean spiral magnetic field we modeled a quasi three-dimensional shock configuration. For the Galaxy, there exist more realistic magnetic field models, like \cite{JF12, SolenoidalJF12} or \cite{TerralFerriere2017}. The magnetic field structure certainly has a great impact on the possibility of CRs travelling back to the Galaxy. Merten et al. \cite{MertenEA2018} showed that the length of the magnetic field lines and the amount of diffusion parallel to the magnetic field lines is critical for a contribution of CRs re-accelerated at the GWTS to the spectrum. The magnetic field also has an impact on the arrival direction of CRs. With CRPropa3.2 the magnetic field can easily be exchanged and the impact on the spectrum and arrival direction on Earth can be studied for various magnetic fields.

\appendix
\section{Advective Step}
\label{sec:AdvectiveStep}

\begin{figure}[htbp] 
        \centering 
        \includegraphics[width=0.8\textwidth]{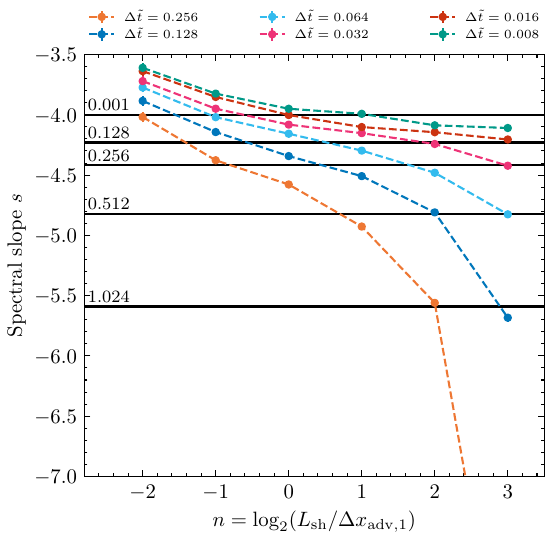}
        \caption{\label{fig:TimeStepShockWidth} Spectral slope of the stationary solution at the shock depending on the time step (indicated by color) and the ratio of upstream advective step and shock width $n$. The expected spectral index for finite shock widths is indicated by horizontal lines (data from \citep{KruellsAchterberg94}).}
         \label{fig:DetailedAdvectiveStep}
    \end{figure}

In Section \ref{sec:1DConstraints} the constraints of the SDE approach are discussed. Figure \ref{fig:DetailedAdvectiveStep} is a detailed version of Fig.\ \ref{fig:AdvectiveDiffusiveStep} where shock width and time step are entangled. Analogously to \citep{KruellsAchterberg94}, for different simulation time steps $\Delta t$ (or advective steps $\Delta x_{\mathrm{adv,1}}$, with $\tilde{u}_1 = 1$) the shock width $L_{\mathrm{sh}}$ is varied. Going along the dotted lines, the shock width increases from left to right. Depending on the shock width, the expected spectral slope according to \citep{KruellsAchterberg94} is indicated. In general, the smaller the time step, the better the predicted spectral slope is met.

\section{Diffusive Step}
\label{sec:AppendixBoundaryDiffusion}

\begin{figure}[htbp] 
        \centering 
        \includegraphics[width=0.8\textwidth]{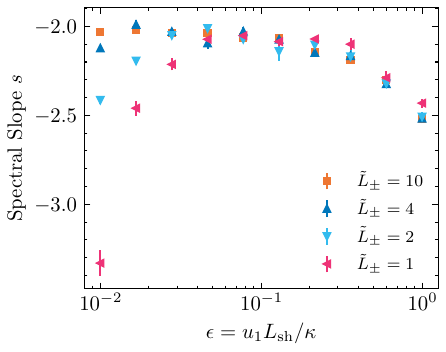}
        \caption{Slope of the stationary spectra at the shock depending on the diffusion coefficient $\kappa$ for different free-escape boundaries $L_{\pm}$. With the diffusion coefficient too large compared to the acceleration region, acceleration is less efficient and the spectral slope decreases. }
        \label{fig:DiffusiveStepBoundary}
    \end{figure}

One advantage of the SDE approach is that there is no need for boundary conditions. However, free escape boundaries can impact the acceleration process and with that the spectral slope. Free escape boundaries are also set in the work of \citep{AchterbergSchure2010}. Figure \ref{fig:DiffusiveStepBoundary} shows how free escape boundaries impact the resulting spectra, depending on $\kappa$ analogous to Fig.\ \ref{fig:AdvectiveDiffusiveStep}. The higher the diffusion coefficient, the better would be the approximation of the ideal shock spectrum. The smaller the acceleration region, defined by $L_{\pm}$, the more particles leave through the free escape boundaries which leads to steeper spectra. 

\section{VLUGR3 Specification}

    We compared the SDE approach with a finite-difference method using VLUGR3 \citep{VLUGI, VLUGII}. The computational domains were chosen to be $x\in \left[ -75x_0, 75x_0 \right]$ and $s = \ln \left( \frac{p}{p_0} \right) \in \left[ -7.5, 15.0 \right]$. The resolution was chosen to be $600 \times 450$ grid points. The time steps were self-determined by VLUGR3 with a given minimal time step of $\delta _{\text{min}}=10^{-9} t_0$.
    The $\delta-$functions for the monoenergetic injection at the shock were approximated by Gaussian functions of the form $\delta _{\text{approx}} \left(  y \right) = \frac{1}{\sqrt{\pi }dy} \exp \left\lbrace - \frac{y^2}{dy ^2} \right\rbrace $. For both $\delta _{\text{approx}} \left(  p \right)$ and $\delta _{\text{approx}} \left(  x \right)$ it was $dp = dx = 0.01$. At last the boundary conditions on the computational domain were given as $\frac{\partial f}{\partial x} = 0$ at $x \pm 75x_0$ and $\frac{\partial f }{\partial s} = $ at $s = -7.5$ v $s = 15.0$, so vanishing gradients on all four boundaries.
    
\section{Approximation of Time-dependent Solution}
\label{sec:FormanApprox}

\begin{figure}
     \centering
     \includegraphics[width = 1\textwidth]{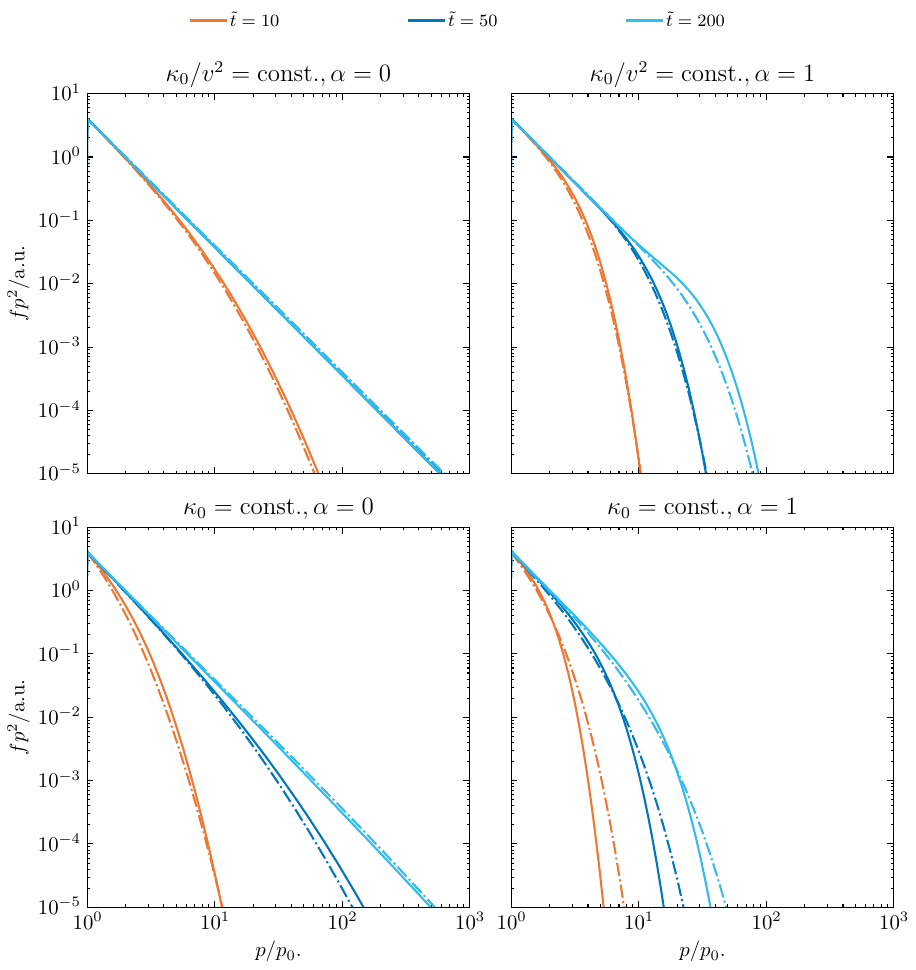}
     \caption{Solution obtained with VLUGR3 (solid line) compared to the approximation of the time-dependent spectrum by \citep{FormanDrury83} (dashed-line) for different diffusion coefficients $\kappa = \kappa_0 (x) \left(p/p_0\right)^{\alpha}$. For $\kappa/u^2 = \mathrm{const.}$ acceleration is so fast that the stationary spectrum is already reached at $\tilde{t} = 50$ for $p < 10^3\,p_0$. The more the diffusion coefficient deviates from $\kappa/u^2 = \mathrm{const.}$, the greater the difference between approximation and numerical solution.}
     \label{fig:FormanApprox}
\end{figure}

In Section \ref{sec:AccTimeScale} we compare the mean acceleration time (e.g. \citep{Drury, FormanDrury83}) with the mean acceleration time calculated from CRPropa3.2 simulations. We use the approximation given by \citep{FormanDrury83} to also compare the shape of the spectrum over time. The approximation is exact in case of $\kappa / u^2 = \mathrm{const.}$, which is the solution derived by \citep{Toptyghin}. Currently, this cannot be handled by CRPropa3.2 and we only compare the solution obtained with the grid-based method VLUGR3. Figure \ref{fig:FormanApprox} shows the time evolution of spectra spatial dependence of the diffusion coefficient and spatially constant diffusion, with both energy-independent, $\alpha = 0$, and energy-dependent diffusion, $\alpha = 1$. The approximation and VLUGR3 solution for constant diffusion is in good agreement. With energy-dependent diffusion our results diverge from the approximation.

\acknowledgments
We acknowledge support from the DFG within the Collaborative Research Center SFB1491 "Cosmic Interacting Matters - From Source to Signal". Additional support from DFG projects EF98/4-1 and FI706/26-1 is acknowledged.


\bibliographystyle{jhep}
\bibliography{references}

\end{document}